\documentclass{article}
\usepackage{graphicx}
\usepackage{float}
\usepackage{dblfloatfix}
\makeatletter
\usepackage[CaptionAfterwards]{fltpage}
\let\saved@includegraphics\includegraphics
\AtBeginDocument{\let\includegraphics\saved@includegraphics}
\renewenvironment*{figure}{\@float{figure}}{\end@float}
\makeatother
\graphicspath{Figures}
\usepackage[margin=1.0in]{geometry}

\title{Dalton's and Amagat's laws fail in gas mixtures with shock propagation}

\author{P. Wayne$^{*}$$^{1}$, S. Cooper$^2$, D. Simons$^{3}$, I. Trueba-Monje$^{4}$,\\ D. Freelong$^{1}$, G. Vigil$^{1}$, P. Vorobieff$^{1}$, C. R. Truman$^{1}$,\\ V. Vorob'ev$^{5}$, \& T. Clark$^{1}$}

\begin{document}
	
	\maketitle
	
	\author[1] Mechanical Engineering Department, University of New Mexico
	\author[2] J. Mike Walker '66 Department of Mechanical Engineering, Texas A\&M University
	\author[3] Department of Aeronautics and Astronautics, Air Force Institute of Technology
	\author[4] Aerospace Engineering Deptartment, Ohio State University
	\author[5] Joint Institute for High Temperatures, Russian Academy of Science

	\begin{abstract}
As a shock wave propagates through a gas mixture, pressure, temperature, and density increase across the shock front. Rankine-Hugoniot (R-H) relations quantify these changes, correlating post-shock quantities with upstream conditions (pre-shock) and incident shock Mach number \cite{White,Zucker,Goldman,Anderson,John}. These equations describe a calorically perfect gas, but deliver a good approximation for real gases, provided the upstream conditions are well-characterized with a thermodynamic mixing model. Two classic thermodynamic models of gas mixtures are Dalton's law of partial pressures and Amagat's law of partial volumes \cite{Wayne}. Here we show that neither thermodynamic model can accurately predict the post-shock quantities of interest (temperature and pressure), on time scales much longer than those associated with the shock front passage, due to their implicit assumptions about behavior on the molecular level, including mixing reversibility. We found that in non-reacting binary mixtures of sulfur hexafluoride (SF$_6$) and helium (He), kinetic molecular theory (KMT) can be used to quantify the discrepancies found between theoretical and experimental values for post-shock pressure and temperature. Our results demonstrate the complexity of analyzing shock wave interaction with two highly disparate gases, while also providing starting points for future theoretical and experimental work and validation of numerical simulations.
	\end{abstract}
\clearpage
	\section*{Introduction}
	In 1802, John Dalton's publication in {\textit{Memoirs of the Literary and Philosophical Society of Manchester}} formulated the law of additive (or partial) pressures~\cite{Dalton}, stating that the total pressure in a non-reactive gas mixture -- at constant temperature and volume -- is equal to the sum of the partial pressures of the component gases.
	
	In 1880, French physicist \'Emile Hilaire Amagat published his findings while researching the compressibility of different gases~\cite{Wayne}. Amagat's law of partial volumes states that the total volume of a gas mixture is equal to the sum of the partial volumes each gas would occupy if it existed alone at the temperature and pressure of the mixture~\cite{Cengel}.  While some advancements have been made in experiments of shock interaction with a single-component gas \cite{Bernstein,Davies}, much less is known about multi-component gas mixtures.
	
	Shock interactions with gas mixtures are relevant to many engineering problems, including gas-cooled reactor power plants~\cite{Tournier,Diaz}, mixing processes in supersonic and hypersonic combustion~\cite{Ferri,Gruenig,Tao,Marble}, and astrophysical phenomena~\cite{Pablo,Anand}. Our experiment was originally designed to determine which thermodynamic law (Dalton or Amagat) is more suitable for predicting properties of gas mixtures interacting with a planar shock wave. Post-shock properties are obtained using the Rankine-Hugoniot equations~\cite{White,Zucker,Goldman,Anderson,John} , which calculate post-shock values, such as pressure, temperature, and density, based on incident shock Mach number and upstream (pre-shock) conditions.

	For a proper comparison of theoretical and experimental values of post-shock properties, pressure and temperature must be measured immediately before (downstream of), and immediately after (upstream of) the shock front. Pressure measurements are not difficult: high frequency response pressure transducers (PTs) are readily available. Temperature measurements are more challenging. Thermocouples are intrusive and lack the necessary response time [$\mathcal{O}(10^{-6})$ sec)]. Infrared (IR) detectors, on the other hand, have ultra-fast response times [$\mathcal{O}(10^{-8}$ sec)], and are inherently non-intrusive. Here we present temperature measurements using an Infrared Associates Mercury-Cadmium-Telluride (MCT), liquid-nitrogen-cooled infrared detector operating at 77 K, with a response time of 60 ns ($6 \times 10^{-8}$ sec). Coupled with a Thorlabs stabilized broadband infrared light source, with a color temperature of 1500 K, the MCT detector provides line-of-sight bulk temperature measurements both before and after the shock. 
	
	For our experiment, we selected two highly disparate gases forming a binary gas mixture: sulfur hexafluoride (SF$_6$) and helium (He). SF$_6$ and He are relatively inexpensive, non-toxic, and have highly contrasting properties: molecular weight, viscosity, specific heat, presenting an extreme (and hopefully easy to interpret) case of a mixture with easily distinguishable components. Two molar concentrations of each gas were chosen; 50\%/50\% (50/50) and 25\%/75\% (25/75) sulfur hexafluoride to helium respectively. 
	
	The shock tube consists of two sections: driver and driven. The driver section is pressurized (with nitrogen) to a value depending on the desired strength of the shock wave. The driven section is pressurized with the test gas mixture, up to one of 3 different initial pressures: 39.3 kPa, 78.6 kPa, or 118 kPa (the average local atmospheric pressure in the lab is approximately 78.6 kPa). A thin-film polyester diaphragm separates the two sections. Once both sections have been evacuated using a vacuum pump, the driver is then filled with the driver gas, and the driven section is filled with our test gas mixture. When the driver and driven sections are at the desired pressure, a pneumatically-driven stainless steel rod, tipped with a broad arrowhead, ruptures the diaphragm, sending a planar shock into the driven section. Four pressure transducers, located on the top of the driven section, record the pressure pulse from the shock wave as it passes. The MCT detector and IR source are located coincident with the 4th downstream pressure transducer, providing nearly instantaneous temperature measurements immediately before, and immediately after the shock. For details of experimental methods and theoretical evaluation of post-shock properties, refer to the Methods and Materials section.
	
\begin{figure}[H]
		\centerline{\includegraphics[width=12cm]{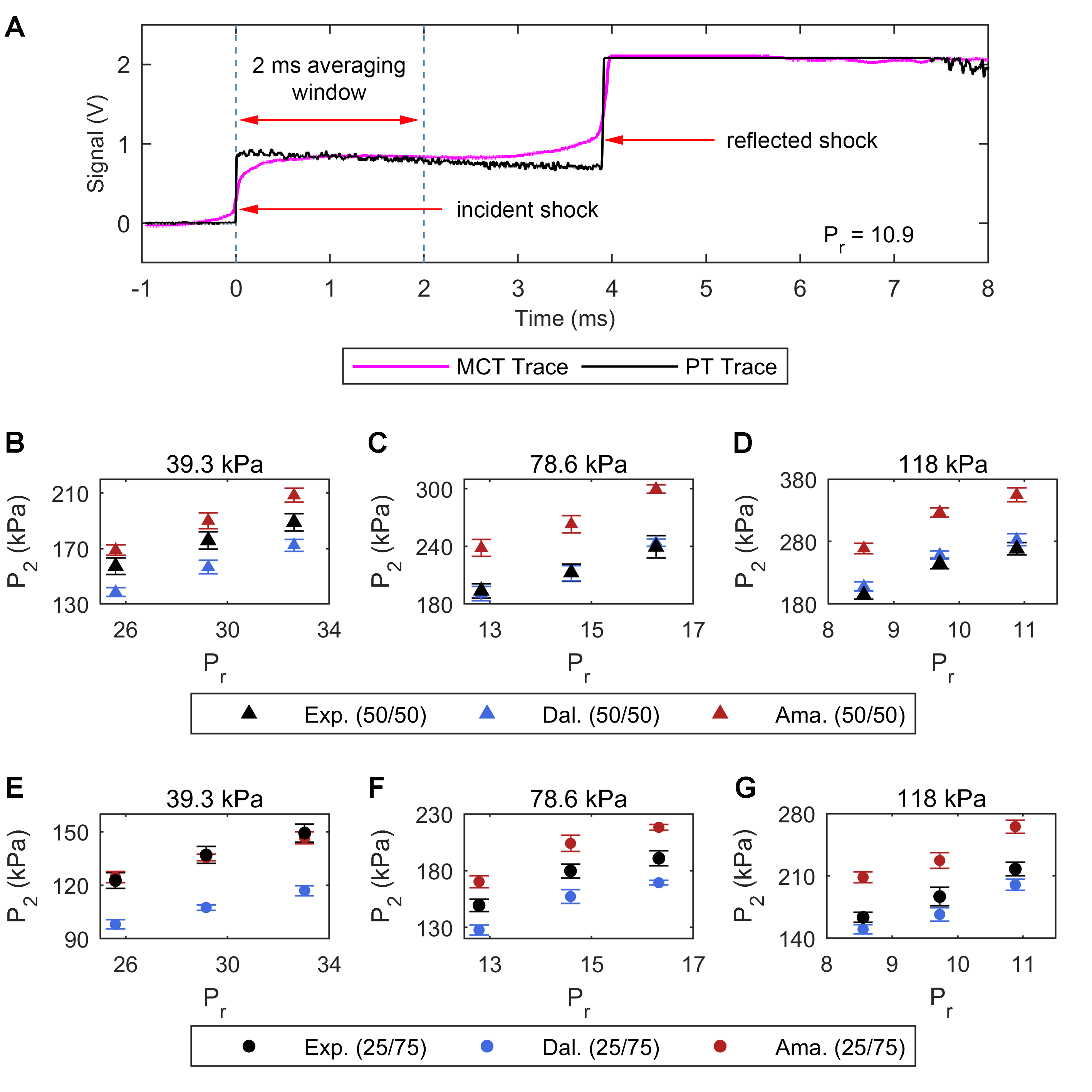}}
		\caption{{{\bf{Time history of recorded signals and post-shock pressure dependence on initial pressure ratio.}} {\bf{A.}} Sample pressure transducer (PT) and MCT detector signal traces, showing locations of incident and reflected shocks, as well as the 2 ms averaging window (dashed blue lines) used in data analysis. The black line is the signal trace from the 4th downstream PT, and the magenta line is the MCT signal trace. {\bf{B.-D.}} Post-shock pressure, $P_2$, versus pressure ratio, $P_r$ for a 50/50 (by mole) binary mixture of SF$_6$ and helium, respectively. {\bf{E.-G.}} Post-shock pressure versus pressure ratio for a 25/75 binary mixture of SF$_6$ and helium. In all panels ({\bf{B.-G.}}), black symbols correspond to experimental values, blue symbols represent Dalton's law predictions, and red symbols denote Amagat's law predictions. Vertical error bars correspond to total uncertainty in post-shock pressure $P_2$. Horizontal error bars are omitted as they do not extend past the physical size of the symbols. The driven pressure associated with each dataset is displayed above the corresponding panel.}}\label{Fig1}
	\end{figure}

	Each gas mixture was tested at three driver pressures (1006 kPa, 1145 kPa, 1282 kPa), and each of the these was applied to three initial pressures (39.3 kPa, 78.6 kPa, 118 kPa) in the driven section, providing experimental datasets at 9 distinct pressure ratios (overpressures $P_r$), where pressure ratio is defined as the ratio of driver pressure to the driven pressure ($P_r = P_{driver}/P_{driven}$). Figure~\ref{Fig1}A is an experimental signal trace at a pressure ratio of $P_r=10.9$, for a 50\%/50\% mixture of SF$_6$/He.

	The black line shows the signal (V from the 4th downstream pressure transducer (PT), the magenta line -- the signal (V) from the MCT detector. Average values of the signals from each PT are determined in a 2 millisecond window after shock impact (dashed blue line). This average value, combined with a calibration curve for each PT (provided by the manufacturer), provides us with the post-shock pressure, $P_2$ (kPa). To obtain the incident shock velocity, $u_1$ (m/s), the distance between two successive pressure transducers is divided by the time it takes the shock wave to travel between them. Measurements of temperature for the MCT are obtained by taking the maximum value of the signal within the same 2 ms window (Fig.~\ref{Fig1}A). A rigorous calibration experiment was conducted with the MCT detector and IR light source to determine post-shock temperature, $T_2$ (K).
	
	\section*{Results}
\begin{figure}[!htbp]
		\centerline{\includegraphics[width=12cm]{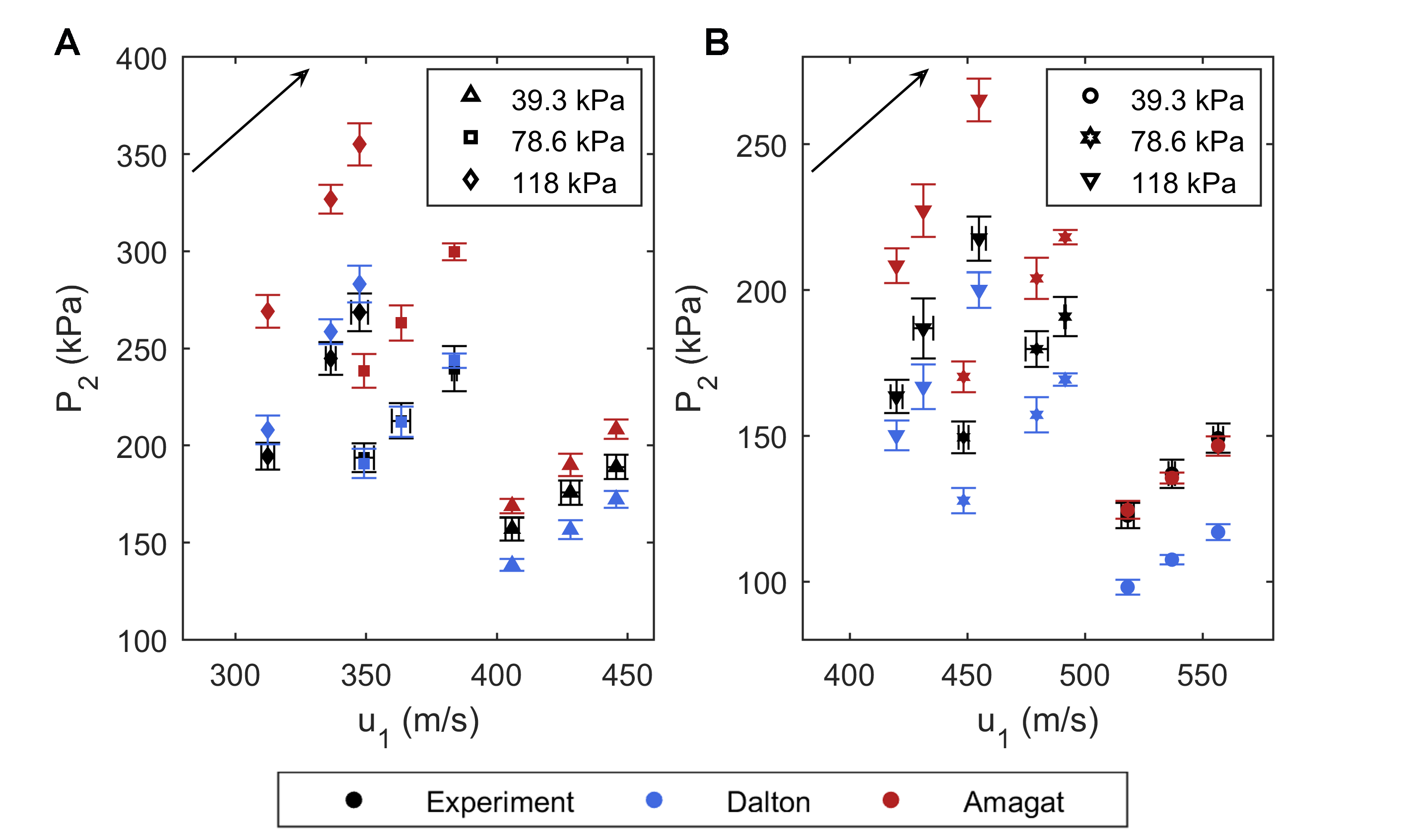}}
		\caption{{{\bf{Post-shock pressure variation with incident shock speed.}} {\bf{A.}} Post-shock pressure ($P_2$) versus incident shock speed ($u_1$) for a 50/50 (by mole) binary mixture of SF$_6$ to helium, respectively. {\bf{B.}} Post-shock pressure versus incident shock speed for a 25/75 binary mixture of SF$_6$ to helium. In both plots, experimental values are given by black symbols, blue symbols correspond to Dalton's law predictions, and red symbols represent Amagat's law predictions. Black arrows show the direction of increasing driver pressure from 1006 kPa to 1282 kPa. Vertical error bars  correspond to the total uncertainty in post-shock pressure $P_2$ (experimental measurements and theoretical predictions), which includes random and systematic uncertainties. Horizontal error bars correspond to total uncertainty in velocity measurements. Horizontal error bars for theoretical predictions are omitted, as the uncertainty in incident shock speed for both thermodynamic laws are identical to that of the experimental measurements.}}\label{Fig2}
	\end{figure}

\begin{figure}[!htbp]
	\centerline{\includegraphics[width=12cm]{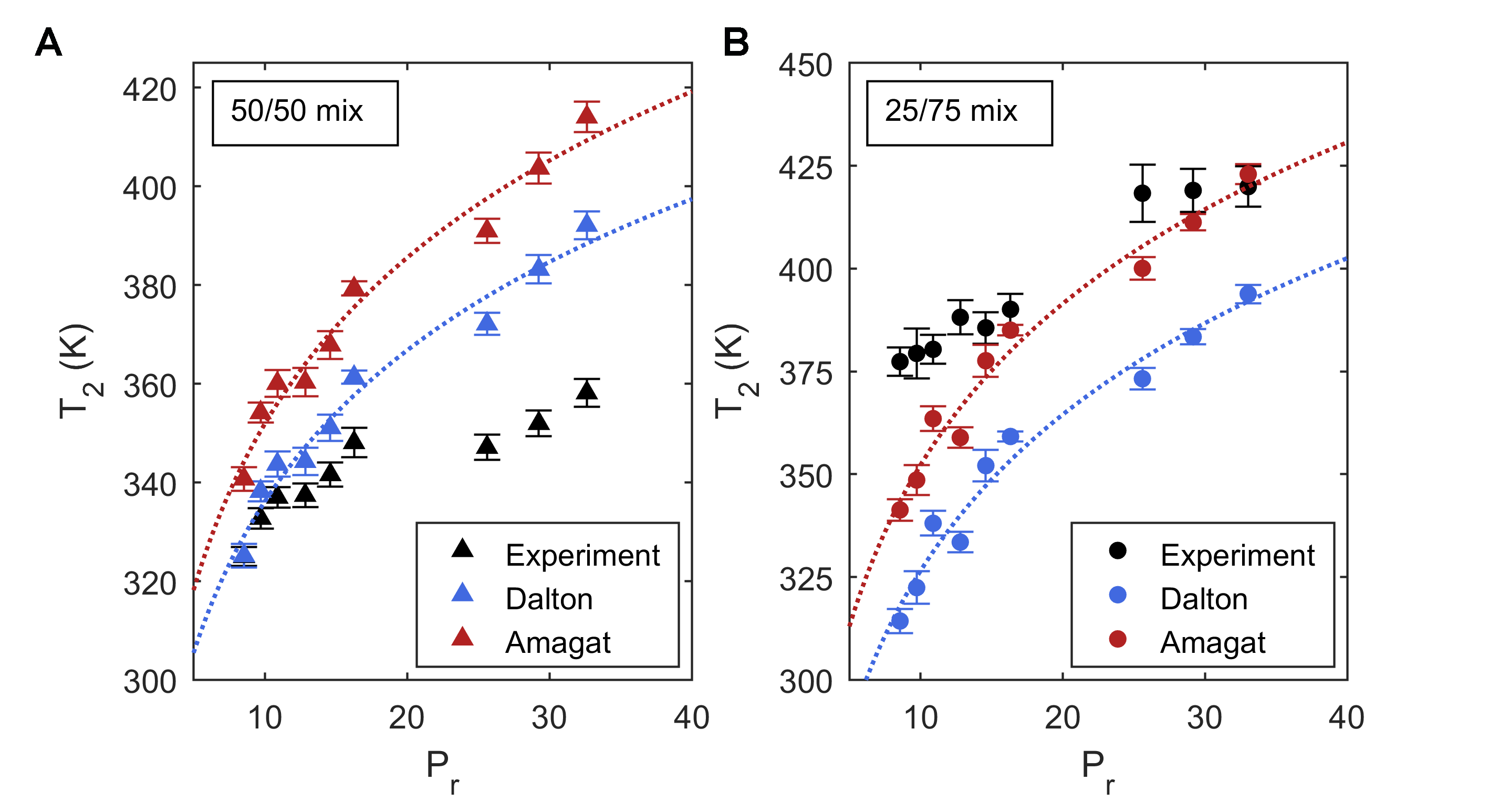}}
	\caption{{{\bf{Post-shock temperature dependence on initial pressure ratio.}} In both plots, black symbols correspond to experimental values, blue symbols represent to Dalton's law predictions, red symbols denote Amagat's law predictions. Blue and red dotted lines in both plots are curve fits to Dalton's and Amagat's law predictions, respectively; they are simply used as guides to the eye. Vertical error bars correspond to total uncertainty in post-shock temperature $T_2$, which includes both random and systematic uncertainties. {\bf{A.}} post-shock temperature ($T_2$) versus pressure ratio ($P_r$) for a 50/50 binary mixture of SF$_6$ to helium. Here, experimental values are closer to Dalton's law predictions. {\bf{B.}} post-shock temperature versus pressure ratio for a 25/75 binary mixture of SF$_6$ to helium. In contrast to data presented in {\bf{A}}, experimental values for a 25/75 mixture are closer to Amagat's law predictions.} }\label{Fig3}
\end{figure}

Multiple experiments were conducted for each gas mixture; a minimum of 6 experiments for each of the 9 pressure ratios. $P_r$ ranges from 8.54 to 32.6 for the 50/50 SF$_6$/He mixture, and from 8.56 to 33.0 for the 25/75 SF$_6$/He mixture. Experimental results in Figs.~\ref{Fig1}-\ref{Fig4} were produced by statistical analysis~\cite{Wheeler} of the dataset of measurements described above and correspond to mean values for each pressure ratio. Theoretical predictions are calculated as follows: the equations of state (EOS) are used to characterize the component gases. An ideal gas EOS is used for helium, while a virial expansion - up to the 4th virial coefficient - is used for sulfur hexafluoride. Inputs for the EOS are the initial pressure ($P_1$) and temperature ($T_1$) in the driven section fo the shock tube. Once the components have been characterized (the process includes calculations of density, specific heat ratio, thermal expansion coefficient, isothermal compressibility, and speed of sound for both He and SF$_6$), we use Dalton's and Amagat's laws to determine thermodynamic coefficients for the gas mixture. Specifically, we are interested in the specific heat ratio - $\gamma$ - and the speed of sound - $a$. These values, combined with the incident shock speed ($u_1$), are used as inputs to the Rankine-Hugoniot equations, which are then used to determine post-shock temperature and pressure. Predictions displayed in Figs.~\ref{Fig1}-\ref{Fig4} correspond to mean values for each pressure ratio. This process, and corresponding statistical analysis, is explained in detail in the Materials and Methods section.
	
Figure~\ref{Fig1}(B-G) are plots of post-shock pressure ($P_2$) versus pressure ratio ($P_r$) for the 50/50 mix (B,C,D) and the 25/75 mix (E,F,G). The corresponding initial pressures (in the driven section) are displayed above each panel (B-G). Vertical error bars correspond to total uncertainty in post-shock pressure. Horizontal error bars corresponding to uncertainty in pressure ratio ($P_r$) do not extend past the physical size of the symbols and are therefore omitted. Note the strong disagreement not only between experimental values and theoretical predictions, but also between the two mixture concentrations. Not only do these results disagree with both Dalton's law and Amagat's law well beyond experimental uncertainty, the disagreement varies with experimental value, leaving no clear answer: which law describes the experiment better overall?

Figure~\ref{Fig2} shows the relationship between post-shock pressure ($P_2$) and incident shock speed ($u_1$) for A. the 50/50 SF$_6$/He mixture and B. the 25/75 SF$_6$/He mixture. In both plots, black symbols denote experimentally measured values, blue symbols correspond Dalton's law predictions, and red symbols represent Amagat's law predictions. Vertical error bars in both plots denote total uncertainty in post-shock pressure. Horizontal error bars (on experimental values) correspond to total uncertainty in incident shock speed. Note that horizontal error bars for both Dalton's and Amagat's laws are withheld, as the uncertainty in incident shock speed for both thermodynamic laws are identical to those for experimental measurements. Black arrows in Figs.~\ref{Fig2}A. and \ref{Fig2}B. (upper left corner of each panel) show the direction of increasing driver pressure, from 1006 kPa to 1282 kPa. While it seems that experimental values are closer to Dalton's law predictions for the 78.6 kPa and 118 kPa data sets, data for 39.3 kPa is inconclusive. Again note the discrepancies between respective initial pressures and between the two mixtures. The next step is to evaluate post-shock temperature and compare with theoretical predictions.
		
Figure~\ref{Fig3}A. is a plot of post-shock temperature, $T_2$, versus pressure ratio for the 50/50 SF$_6$/He mixture and Fig.~\ref{Fig3}B. is the same plot for the 25/75 SF$_6$/He mixture. Vertical error bars for experimental measurements and theoretical predictions correspond to total uncertainty in post-shock temperature. Note the inputs used to calculate predictions for both Dalton's and Amagat's laws are the initial pressure ($P_1$) and temperature ($T_1$) in the driven section of the shock tube, which are measured values. Therefore, these predictions will not fit smoothly on a curve; the curve fits shown in Fig.~\ref{Fig3}A. and \ref{Fig3}B. serve merely as visual aids.

Again there is strong disagreement between experimentally obtained values of post-shock temperature and theoretical predictions. For the 50/50 mixture, it seems experimental values are closer to Dalton's law predictions, and for the 25/75 mixture, experimental values are closer to Amagat's law predictions; these discrepancies are not random. What could be responsible for this systematic disagreement?

\section*{Discussion}

A compelling theoretical analysis of finite-strength shock propagation through a binary gas mixture was published by Sherman{~\cite{Sherman}} (1960) for inert gas mixtures consisting of argon (Ar) and helium (for a range of molar concentrations of each component). Using a continuum approach, Sherman focused on determining the structure of a shock wave of arbitrary strength, which includes ordinary-, baro-, and thermal-diffusion effects. He concluded that baro-diffusion speeds up the heavier component, and slows down the lighter component relative to the mass velocity of the mixture{~\cite{Sherman}}. He also determined thermal-diffusion will have the opposite effect, slowing down the heavier component and speeding up the lighter. Therefore, thermal diffusion would (at least partially) counteract baro-diffusion within the shock wave. Sherman calculated his results through numerical integration of the Navier-Stokes equations. A primary assumption associated with this analysis was in always imposing thermal equilibrium between species through the shock wave. However, Sherman indicates these assumptions (and corresponding analysis) may not be valid for strong shock waves, or for mixtures with large molecular mass ratios. In fact, he states that one might intuitively expect that the maximum shock strength for which these calculations give realistic predictions would be reduced (somewhat significantly) as the molecular mass ratio increases, due to the difficulty in maintaining thermal equilibrium between the gas components{~\cite{Sherman}}.

In 1967, Bird{~\cite{Bird}} produced an interesting attempt to model shock propagation through a binary mixture of 50/50 Ar-He, representing the gas molecules as rigid elastic spheres - with the appropriate masses and diameters - and compared  his results with the analytical predictions of Sherman. His model did not have the same temperature constraints as Sherman, and predicted even greater differences in velocity, temperature, and concentration profiles than the analytical profiles of Sherman. Furthermore, Bird's findings suggest the temperature non-equilibrium between species increases with low concentrations of the heavy gas component and that this [non-equilibrium] can persist for a considerable distance downstream of the shock{~\cite{Bird}}. For reference, the molecular mass ratio with respect to binary mixtures of Ar-He is approximately 10, while the molecular mass ratio associated with the current work (SF$_6$ - He) is around 36.5.

The studies conducted by Sherman and Bird reasonably concluded that shock propagation through gas mixtures with relatively massive molecules would cause differences in their behavior, but the results only pertained to gas molecule velocity, concentration, and temperature profiles - on time (and length) scales much smaller than these considered in the current work. Perhaps the results shown in Figs.~\ref{Fig1}-\ref{Fig3}, taken in context with these early studies, suggest a different explanation, beyond experimental uncertainty.

Is a kinetic molecular theory (KMT){~\cite{Loeb,Collie}} explanation possible? Dalton's and Amagat's laws predate KMT, but each law makes implicit assumptions about reversibility. Both laws assume thermodynamic equilibrium. However, while Dalton's law assumes the gases are always perfectly mixed, Amagat's law assumes the gases will separate over time. This assertion is interesting because that is exactly what we see in experiments. The molecular mass of sulfur hexafluoride is approximately 36.5 times greater than that of helium. If the gases are not constantly mixed, they will begin to separate (under normal conditions, the gases will separate within a few minutes). Prior to taking measurements described here, our initial guess was that Amagat's law might provide a better prediction for these dissimilar gases. The time scale of that separation, however, is much longer than the time scale of the experiment associated with the shock passage, thus leading to both Dalton's law (infinite separation time) and Amagat's law (the gases in the mixture are in effect always separated) failing to produce a match with experiment.

What relaxation time scale is relevant for the equilibrium status of a shocked gas mixture? Relaxation time is defined as the time within which a perturbed gas will reach statistical (thermodynamic) equilibrium~\cite{Landau}. In binary gas mixtures whose constituents have widely different molecular masses ($M_{SF_{6}}$ $\gg$ $M_{He}$), disparate relaxation times are manifest, governing the approach to equilibrium of the various degrees of freedom~\cite{Mora}. According to Mora and Fern\'{a}ndez-Feria~\cite{Mora}, for such mixtures, the process of equilibration can be characterized by three different relaxation times: two for self-equilibration of the component gases, and a third one associated with the slower process of interspecies equilibration. For this analysis, these relaxation times are the post-shock mean free time (or average time between molecular collisions) for each of the gas components ($\tau_{He}$ and $\tau_{SF_{6}}$) and the change in collision time between them ($\Delta{\tau}=\tau_{He} - \tau_{SF_{6}}$). What is remarkable about this method is that the mean free time is temperature and pressure dependent (see Methods). 

\begin{figure}[!htbp]
	\centerline{\includegraphics[width=12cm]{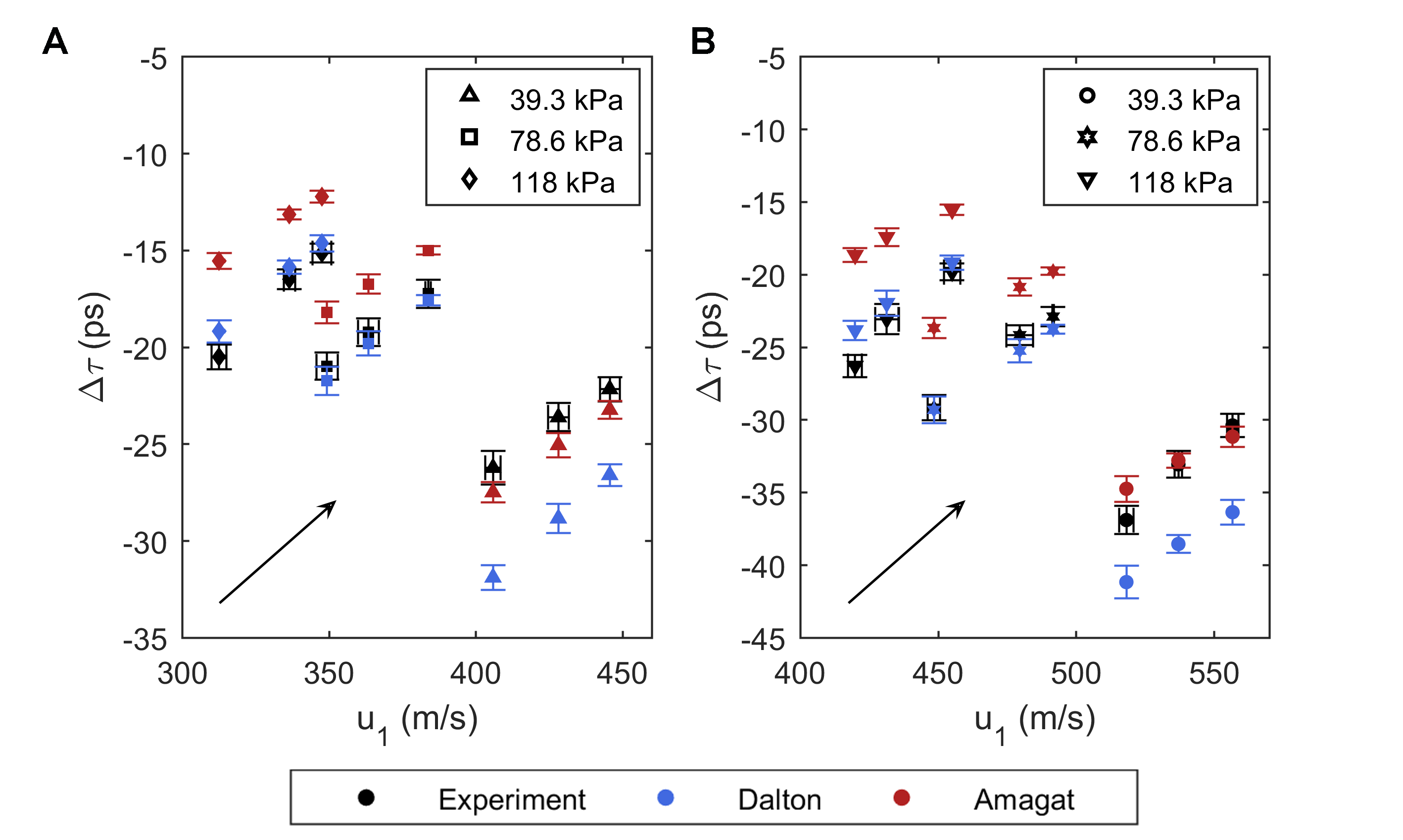}}
	\caption{{{\bf{Kinetic theory correlations with incident shock speed.}}{\bf{A.}} change in collision time, $\Delta{\tau} = {\tau}_{He} - {\tau}_{SF_6}$, for a 50/50 (by mole) binary mixture of SF$_6$ to helium, respectively. {\bf{B.}} change in collision time for a 25/75 binary mixture of SF$_6$ to helium. In both plots, experimental values are given by black symbols, blue symbols denote Dalton's law predictions, and red symbols represent Amagat's law predictions. Vertical error bars correspond to total uncertainty in $\Delta{\tau}$, and horizontal error bars correspond to total uncertainty in velocity measurements. Similar to Fig.~\ref{Fig2}, horizontal error bars for theoretical predictions are omitted}, and black arrows show the direction of increasing driver pressure from 1006 kPa to 1282 kPa. These data closely resemble the trends we observe above (Figs. 1-3).}\label{Fig4}
\end{figure}

Figure~\ref{Fig4} is a plot of the change in collision time, $\Delta{\tau}$ (in picoseconds), versus incident shock velocity ($u_1$) for A. the 50/50 SF$_6$/He mixture and B. the 25/75 SF$_6$/He mixture. Symbol type and color in Fig.~\ref{Fig4} have been arranged similar to Fig.~\ref{Fig2}. Again, the black arrow in both plots points in the direction of increasing driver pressures. Note again the systematic discrepancies in these data, consistent with the spread of experimental data points. The negative values for $\Delta{\tau}$ are due to the fact that the average collision time for SF$_6$ is an order of magnitude greater than that of He. While Figs.~\ref{Fig1}-\ref{Fig3} show discrepancies produced by Dalton's and Amagat's laws in predicting post-shock properties (pressure and temperature) on a macroscopic scale, Fig.~\ref{Fig4} provides a context for these discrepancies, relating them to the simplest quantitative parameter describing the disparity between component gas behavior on a microscopic scale. Therefore, if the component gases behave differently, i.e. have a large difference in response time, on a microscale, is it not reasonable to assume these discrepancies manifest on a macroscale? This hypothesis is reinforced by the results obtained from Sherman{~\cite{Sherman}} and Bird{~\cite{Bird}}, especially considering the large molecular mass difference between the species.

The simple explanation that kinetic molecular theory can provide is that differences in the response time of the molecules account for the disagreement between theory and experiment. This explanation appears to agree with our data, at least qualitatively. These observations show that Dalton's and Amagat's laws fail to accurately describe the behavior of a gas mixture that underwent shock acceleration, with implications that the same failure can manifest in other non-equilibrium situations.
		\section*{Materials and Methods}	
		
		\subsection*{Experimental setup}
		
		Two gas mixtures were tested: 50\%/50\% and 25\%/75\% sulfur hexafluoride (SF$_6$) to helium (He), by mole, respectively. Experiments were conducted at the Shock Tube Facility in the Mechanical Engineering Department at the University of New Mexico (UNM). The shock tube itself is approximately 5.2 m long, with a 2 m long driver section, and a 3.2 m long driven section (7.62 cm inside square cross-section). 
		
		To begin an experiment, we separate the driver and driven sections of the shock tube with a thin-film polyester diaphragm. Both sections are then evacuated using an ITT Pneumotive vacuum pump to -78.6 kPa. The driven (test) section is filled with the test mixture to a predetermined pressure (39.3 kPa, 78.6 kPa, or 118 kPa). The driver is filled with nitrogen to another predetermined pressure (1006 kPa, 1145 kPa, or 1282 kPa). The pressure in the driver section depends on the desired strength (or Mach number) of the shock wave \cite{Wayne}. Once the pressure in both sections has stabilized, a pneumatically-driven stainless steel rod tipped with a broad arrowhead punctures the diaphragm, sending a normal shock down the length of the driven section. Four pressure transducers (with a response time $\leq$ 1 $\mu$s), located on the top of the driven section ($\approx$ 0.8 meters apart), record the pressure history of the shock wave as it passes. These data can then be used to determine the velocity of the shock wave, $u_1$, and the post-shock pressure, $P_2$. The MCT detector and IR source are placed on opposite sides of the shock tube, perpendicular to the optical axis and located coincident with the 4th downstream pressure transducer. Two zinc selenide (ZnSe) optical windows are placed on either side of the shock tube, mounted flush with the inside of the test section. These optical windows are used to create an air-tight, unobstructed light path from the IR source, through the test gas in the driven section, to the sensor on the MCT detector. 
		They are also used as broadband filters to decrease the wavelength range of the incoming light from the IR source to between 7 $\mu$m and 12 $\mu$m (the IR source outputs light between 0.5 $\mu$m and approximately 20 $\mu$m). A germanium notch filter, mounted on the MCT detector itself, further reduces the range of incoming light to between 7.5 $\mu$m and 9.0 $\mu$m. Note the target range for these experiments was chosen between 7.5 $\mu$m and 8.5$\mu$m, based on the infrared absorption spectrum of sulfur hexafluoride~\cite{Lagemann}.
		
		Calibration curves that relate the signal (V) from the pressure transducers to pressure (kPa) in the driven section are provided for each transducer by the manufacturer. For the MCT detector, a calibration experiment was conducted to determine the relationship between signal (V) and temperature (K). This experiment used an aluminum cylinder with ZnSe optical windows mounted on each side, with components and geometry identical to that of the shock tube setup described above. The cylinder chamber is evacuated with a vacuum pump and filled with the test gas mixture at a prescribed pressure (target pressures were determined from previous experiments~\cite{Iggy}). A helical coil resistance heater, placed on the inside of the calibration cylinder, is then activated to increase the temperature of the test gas to a predetermined value. A Vincent Associates Uniblitz LS6 laser shutter (6 mm aperture, 1.7 ms open time) is placed along the optical axis in front of the MCT detector, which effectively simulates an instantaneous increase in temperature, as would be seen by the sensor when the shock wave passes~\cite{Wayne}. Once the laser shutter is activated, the signal from the detector is recorded and used as a baseline for that pressure-temperature (P-T) combination. For each gas mixture, data was obtained at up to 25 P-T combinations, with a minimum of six measurements at each combination. This method provided ample data to determine a calibration curve for each gas mixture.
		
		\subsection*{Thermodynamic models}
		
		The following theoretical analysis was used to determine post-shock pressure and temperature for both Dalton's and Amagat's laws. Equations of state (EOS) are needed to characterize the gas components; the Ideal Gas (Eqn. 1) EOS was used for helium and a virial expansion (Eqn. 2) was used for sulfur hexafluoride.
		\begin{eqnarray}
		{P\nu=RT}\label{Ideal} \\
		{P = \frac{RT}{\nu}\left(1 + \frac{B(T)}{\nu} + \frac{C(T)}{\nu^2} + \frac{D(T)}{\nu^3} + \frac{E(T)}{\nu^4}\right)}\label{Virial}
		\end{eqnarray}
		Here, $P$ is the pressure (N/m$^2$), $\nu$ is the specific volume (m$^3$/mol), $R = 8.314$ J/mol-K is the universal gas constant, $T$ is the temperature (K), $B(T)$ is the second virial coefficient, $C(T)$ is the third virial coefficient, and so forth~\cite{Callen}. Algebraic expressions of the temperature provided by the hard-core square well (HCSW) model intermolecular potential are used to represent $B(T)$ and $C(T)$, while $D(T)$ and $E(T)$ are represented as polynomial functions of the inverse temperature \cite{Hurly}. 
		\begin{eqnarray}
		B(T) = b_0[1-(\lambda^3 -1)\Delta]\\
		C(T) = \frac{1}{8}b_0^2(5-c_1\Delta - c_2\Delta^2 - c_3\Delta^3)\\
		D(T) = \sum_{n=0}^3 d_nT^{-n}\\
		E(T) = \sum_{n=0}^3 e_nT^{-n}
		\end{eqnarray}
		where $\Delta = e^{\epsilon/k_b T} - 1$ ($k_b$ is Boltzmann's constant) and the coefficients $c_1$, $c_2$, and $c_3$ are given by
		\begin{eqnarray}
		c_1 = \lambda^6 - 18\lambda^4 + 32\lambda^3 -15\\
		c_2 = 2\lambda^6 - 36\lambda^4 + 32\lambda^3 +18\lambda^2 -16\\
		c_3 = 6\lambda^6 - 18\lambda^4 + 18\lambda^2 -6
		\end{eqnarray}
		Values for $d_n$ ($n=1, 2, 3$), $e_n$ ($n=1,2,3$), $b_0$, $\lambda$, and $\epsilon/k_b$ are provided by Hurly et al \cite{Hurly}. 
		
		Inputs for the EOS are the initial pressure ($P_1$) and temperature ($T_1$) of the test gas in the driven section. Once the component gases have been characterized, we use Dalton's law and Amagat's law to determine thermodynamic coefficients for the mixture. We are specifically looking for the speed of sound $a$, and the specific heat ratio, $\gamma$. These variables, coupled with the incident shock speed $u_1$ (determined from experiment) are used as inputs for the Rankine-Hugoniot equations (Eqns.10-13), which relate post-shock properties in terms of initial conditions (pre-shock) and incident shock Mach number $M_1$.
		\begin{eqnarray}
		{M_1 = \frac{u_1}{a}}\label{M1}\\
		{M_2^2 = \frac{M_1^2 + 2/\left({\gamma}-1\right)}{\left[2{\gamma}/\left({\gamma}-1\right)\right]M_1^2 - 1}}\\
		{T_1\left(1+\frac{\gamma - 1}{2}M_1^2\right) = T_2\left(1+\frac{\gamma - 1}{2}M_2^2\right)}\\
		{\frac{P_1}{P_2} = \frac{1+{\gamma}M_2^2}{1+{\gamma}M_1^2}}\label{P1P2}
		\end{eqnarray}
		The subscripts 1 and 2 correspond to conditions before and after the shock, respectively. Once the post-shock temperature ($T_2$) and pressure ($P_2$) for each thermodynamic law (Dalton and Amagat) have been calculated, we can directly compare the results with experimental values. All theoretical calculations and analysis of experimental data (including statistical analysis) were performed in MATLAB.
		
		\subsection*{Kinetic molecular theory}
		Central to kinetic molecular theory are the following assumptions~\cite{Tro}:
		\begin{itemize}
			\item The size of the particle is negligibly small, i.e. the particles themselves occupy no volume, even though they have mass. At the maximum concentration of molecules in our experiments, the Van der Waals correction to pressure associated with molecular volume would not exceed 2.5\%.\
			\item The average kinetic energy of a particle is proportional to the temperature (K).
			\item Particle collisions are perfectly elastic; they may exchange energy, but there is {\textit{no overall loss of energy}}.
		\end{itemize}
		
		The following equations were used for kinetic theory analysis. The mean free path ($l$) is given by
		\begin{equation}\label{mfp}
		l = \frac{k_bT}{\sqrt{2}{\pi}{d^2}P}
		\end{equation}
		where $k_b \approx 1.381 \times 10^{-23}$ m$^2$ kg s$^{-2}$ K$^{-1}$ is the Boltzmann constant, $T$ is the temperature (K), $d$ is the kinetic diameter, which for helium is 0.260$\times 10^{-9}$ m and for SF$_6$ is 0.550$\times 10^{-9}$ m, and $P$ is the pressure (Pa). 
		
		Mean molecular speed ($\mu_m$) is obtained via
		\begin{equation}\label{mms}
		\mu_m = \sqrt{\frac{8TR_s}{\pi}}
		\end{equation}
		where $R_s$ is specific gas constant: for helium, $R_s = 2.0773\times 10^{-3}$ J/kg-K, and for SF$_6$, $R_s = 56.9269$ J/kg-K.
		
		Average collision time ($\tau$) is calculated using
		\begin{equation}\label{act}
		\tau = \frac{l}{\mu_m}
		\end{equation}
		
		The change in average collision time, $\Delta{\tau}$ is simply the average collision time of helium, $\tau_{He}$, minus the average collision time of sulfur hexafluoride, $\tau_{SF_{6}}$
		\begin{equation}
		{\Delta{\tau} = \tau_{He} - \tau_{SF_{6}}}
		\end{equation}
		
		\subsection*{Statistical analysis} A comprehensive statistical analysis was performed on all experimental data, according to steps outlined in Wheeler and Ganji \cite{Wheeler}.
		
		Statistical analysis on all measurements (pressure, temperature, and velocity) begins with outlier rejection using the Modified Thompson Tau technique \cite{Wheeler}. In this method, for any $n$ measurements, with a mean value $\bar{x}$ and standard deviation $S$, the data is arranged in ascending order ($x_1$, $x_2$, ...,$x_n$). The extreme (highest and lowest) values are suspected outliers. For these suspected points, a deviation is calculated as
		\begin{equation}
		\delta_i = \left|x_i - \bar{x}\right|
		\end{equation}
		and the largest value is selected. This value is compared with the product of $\tau$ (tabulated with respect to $n$) times the standard deviation $S$. If the value of $\delta$ exceeds $\tau S$, then this value can be rejected as an outlier (only one value is eliminated for each iteration). The mean and standard deviation of the remaining values are then recomputed and the process is repeated until no more outliers exist. Note that $n$ decreases with each outlier rejection.
		
		Pressure measurements were obtained using multiple devices. Therefore, an estimation of the combined degrees of freedom according to the Welch-Satterthwaite formula (Eqn. 18) is necessary \cite{Wheeler}.
		\begin{equation}\label{WS}
		\nu_x = \frac{\left[{\sum\limits_{i=1}^m} S_i^2\right]^2}{\sum\limits_{i=1}^m (S_i^4/
			\nu_i)}
		\end{equation}
		where $\nu_i$ is the degrees of freedom for the measuring device, and $\nu_x$ is the value of the combined degrees of freedom for variable $x$. Degrees of freedom for temperature and velocity measurements is simply $\nu_x = n-1$. When $\nu_x$ has been determined, the students t-distribution value ($t$) is found based on a 95\% level of confidence. The total random uncertainty in the mean value $P_{\bar{x}}$ is computed by
		\begin{equation}
		P_{\bar{x}}=\pm \frac{S_x}{\sqrt{n}}
		\end{equation}
		where $S_x/\sqrt{n}$ is the estimate of the standard deviation of the mean. 
		
		Systematic uncertainty for each variable is determined using the mean value for the measurement and manufacturer supplied information, such as linearity, hysteresis, and uncertainty in the measuring device. Sources of systematic error are pressure transducers, pressure gauges, oscilloscopes, and the MCT detector. Total systematic error is given by
		\begin{equation}
		B_x = \left[\sum\limits_{i=1}^k B_i^2\right]^{1/2}
		\end{equation}
		where $B_i$ is the systematic error for measuring device $i$. Once all sources of random and systematic error have been determined, the total uncertainty in the mean ($W_{\bar{x}}$) is given by
		\begin{equation}
		W_{\bar{x}} = \left(B_x^2 + P_{\bar{x}}^2\right)^{1/2}
		\end{equation}
		The mean value for a given set of measurements is used in all plots, and error bars represent total uncertainty ($\pm$) in the mean value. 
		
		Table~\ref{Table1} details the number of measurements $n$, degrees of freedom $\nu$ (combined or otherwise), and the $t$-distribution value for post-shock pressure $P_2$, incident shock speed $u_1$, and post-shock temperature $T_2$, for the 50/50 binary mixture of SF$_6$ and helium. $P_r$ denotes the pressure ratio according to mean values of driver and driven pressures for a given set of experiments. Table~\ref{Table2} details statistics for a 25/75 mixture of SF$_6$ and helium. All statistical analysis was performed in MATLAB.

		\setcounter{table}{0}
		\begin{table}[H]
			\linespread{1.0}\caption{Statistical information of experimental measurements for a 50/50 binary mixture of SF$_6$ and helium, where $n$ is the number of samples, $\nu$ is the degrees of freedom (combined or otherwise), and $t$ is the student's $t$-distribution value.}
			\renewcommand\arraystretch{1.5}
			\begin{center}\label{Table1}
				\begin{tabular}{|c|c|c|c|c|c|c|c|c|c|}
					\hline 
					$P_r$ $(50/50)$&$n_{P_2}$  &$\nu_{P_2}$  &$t_{P_2}$  &$n_{u_1}$  &$\nu_{u_1}$  &$t_{u_1}$  &$n_{T_2}$  &$\nu_{T_2}$  &$t_{T_2}$  \\ 
					\hline 
					32.61&15  &6  &2.4469  &5  &4  &2.7764  &5  &4  &2.7764  \\ 
					\hline 
					16.27&23  &10  &2.2281  &4  &3  &3.1824  &7  &6  &2.4469  \\ 
					\hline 
					10.88&20  &10  &2.2281  &6  &5  &2.5706  &5  &4  &2.7764  \\ 
					\hline 
					29.24&20  &16  &2.1199  &5  &4  &2.7764  &6  &5  &2.5706  \\ 
					\hline 
					14.60&21  &11  &2.2010  &5  &4  &2.7764  &5  &4  &2.7764  \\ 
					\hline 
					9.70& 21 & 10 &2.2281  &5  &4  &2.7764  &6  &5  &2.5706  \\ 
					\hline 
					25.60&22  &15  &2.1314  &6  &5  &2.5706  &6  &5  &2.5706  \\ 
					\hline 
					12.83&22  &15  &2.1314  &5  &4  &2.7764  &6  &5  &2.5706  \\ 
					\hline 
					8.54&22  &17  &2.1098  &6  &5  &2.5706  &6  &5  &2.5706  \\ 
					\hline 
				\end{tabular}
			\end{center}
		\end{table} 
		
		\begin{table}[H]
			\linespread{1.0}\caption{Statistical information of experimental measurements for a 25/75 binary mixture of SF$_6$ and helium, where $n$ is the number of samples, $\nu$ is the degrees of freedom (combined or otherwise), and $t$ is the student's $t$-distribution value.}
			\renewcommand\arraystretch{1.5}
			\begin{center}\label{Table2}
				\begin{tabular}{|c|c|c|c|c|c|c|c|c|c|}
					\hline 
					$P_r$ $(25/75)$&$n_{P_2}$  &$\nu_{P_2}$  &$t_{P_2}$  &$n_{u_1}$  &$\nu_{u_1}$  &$t_{u_1}$  &$n_{T_2}$  &$\nu_{T_2}$  &$t_{T_2}$  \\ 
					\hline 
					33.03&20  &12  &2.1788  &6  &5  &2.5706  &6  &5  &2.5706  \\ 
					\hline 
					16.32&24  &20  &2.0860  &4  &3  &3.1824  &5  &4  &2.7764  \\ 
					\hline 
					10.88&18  &8  &2.3060  &6  &5  &2.5706  &6  &5  &2.5706  \\ 
					\hline 
					29.16&28  &21  &2.0796  &6  &5  &2.5706  &6  &5  &2.5706  \\ 
					\hline 
					14.59&19  &10  &2.2281  &6  &5  &2.5706  &6  &5  &2.5706  \\ 
					\hline 
					9.73&21  &12  &2.1788  &6  &5  &2.5706  &6  &5  &2.5706  \\ 
					\hline 
					25.59&21  &11  &2.2010  &6  &5  &2.5706  &6  &5  &2.5706  \\ 
					\hline 
					12.79&20  &13  &2.1604  &4  &3  &3.1824  &6  &5  &2.5706  \\ 
					\hline 
					8.56&18  &10  &2.2281  &6  &5  &2.5706  &6  &5  &2.5706  \\ 
					\hline 
				\end{tabular}
			\end{center}
		\end{table}

		\subsection*{Acknowledgments} We would like to thank Josiah Bigelow for his comments, suggestions, and numerical work. 
		
		\subsection*{Funding} We acknowledge support from the National Nuclear Security Administration (NNSA) grant DE-NA-0002913. 
		
		\subsection*{Author contributions} All authors contributed equally to this work. Authors (1-6) supported by NNSA grant DE-NA-0002913. 
		
		\subsection*{Competing interests} The authors declare that they have no competing financial interests. 
		
		\subsection*{Data and materials availability} All data and corresponding Matlab scripts used to support the findings in this study are available upon request from the corresponding author. Requests for materials should be addressed to Patrick Wayne (email: patrick.j.wayne@gmail.com) or Peter Vorobieff (email: kalmoth@unm.edu).
		\\
		\\
	
		\clearpage

\end{document}